\documentclass[submission]{eptcs}
\usepackage{xspace}
\usepackage{graphicx} 
\usepackage{amsmath}
\usepackage{amssymb}
\usepackage[usenames]{color}

\definecolor{MyDarkBlue}{rgb}{0.1,0,0.55}
\definecolor{MyDarkRed}{rgb}{0.55,0,0.1}

\newcommand{\elide}[1]{}

\newcommand{\mref}[1]{(\ref{eq:#1})}
\newcommand{\Proof}{\noindent\textbf{Proof:}\xspace}

\newcommand{\EndProof}{\xspace\textbf{EndProof.}}

\newcommand{\abs}[1]{\overline{#1}}

\newcommand{\outline}[1]{{\color{MyDarkBlue}{#1}}}

\newtheorem{thm}{Theorem}
\newtheorem{lem}{Lemma}
\newtheorem{cor}{Corollary}
\newtheorem{obs}{Observation}

\newtheorem{defn}{Definition}

\newcommand{\theorem}[1]{ \begin{thm} #1 \end{thm}}

\newcommand{\st}{\;|\;}

\newcommand{\false}{\mathit{false}}
\newcommand{\true}{\mathit{true}}

\newcommand{\Equiv}{\;\equiv\;}
\newcommand{\Implies}{\;\Rightarrow\;}
\newcommand{\AAnd}{\;\wedge\;}
\newcommand{\Or}{\;\vee\;}
\newcommand{\Not}{\neg}

\newcommand{\Union}{\cup}

\newcommand{\Diff}{\backslash}

\newcommand{\becomes}{\;:=\;}

\newcommand{\unch}{\mathit{unch}}    

\newcommand{\SP}{\mathit{SP}}        

\newcommand{\pA}{\mathsf{A}}

\newcommand{\guards}{\;\longrightarrow\;}

\newcommand{\Reach}{\mathit{Reach}}
\newcommand{\comment}[1]{}

\newcommand{\dia}[1]{\langle{#1}\rangle}

\title{Model Checking in Bits and Pieces}
\author{
  Kedar S. Namjoshi
  \institute{Bell Labs, Alcatel-Lucent}
  \email{kedar@research.bell-labs.com}
}
\def\titlerunning{Model Checking in Bits and Pieces}
\def\authorrunning{K.S. Namjoshi}

\begin{document}
\maketitle

\begin{abstract}

Fully automated verification of concurrent programs is a difficult problem, 
primarily because of \emph{state explosion}: the exponential growth of a program state
space with the number of its concurrently active components. It is natural to apply a 
divide and conquer strategy to ameliorate state explosion,  by 
 analyzing only a single component at a time. We show that this strategy leads to the notion of a  ``split"
invariant, an assertion which is globally inductive, while being structured as the conjunction of a number of local, per-component invariants. This formulation is closely connected to the classical Owicki-Gries method and to Rely-Guarantee reasoning. We show how the division of an invariant into a number of pieces with limited scope makes it possible to apply new, localized forms of symmetry and abstraction to drastically simplify its computation.  Split invariance also has interesting connections to parametric verification. A quantified invariant for a parametric system is a split invariant for every instance. We show how it is possible, in some cases, to invert this connection, and to automatically generalize from a split invariant for a small instance of a system to a quantified invariant which holds for the entire family of instances.

\end{abstract}

\section{Introduction}
\comment{
\outline{
-- Concurrency, which was once limited to operating systems and networking, is now getting into the mainstream, due to multicores -- Verification of conc. programs is hard, as a proof has to simultaneously track the behavior of multiple executing threads -- Theoretically, it is PSPACE hard (give proof) -- Strategy is divide and conquer -- pros and cons -- sketch of paper.
}}

Concurrency was once limited to the internals of operating systems and
networking. It is now in the mainstream of programming, largely due to the
availability of cheap hardware and the ubiquitous presence of multi-core
processors. Designing a correct concurrent program or protocol, however, is a
hard problem. Intuitively, this is because the designer must coordinate the behavior of multiple, simultaneously active threads of execution. Verification of an existing program is an even harder problem, as the analysis process must reconstruct the invariants which guided the original design of the program. These informal statements can be made precise through complexity theory: verification of an $N$-process concurrent program is PSPACE-hard in $N$, even if the state space of each component is fixed to a small constant. In practice, the difficulty manifests itself as model checking tools run into \emph{state explosion}: the exponential growth of a program state space with the number of its concurrent components. 

A common strategy when faced with a large problem is to break it into smaller and simpler sub-problems. 
In program verification, this ``divide and conquer" strategy is known as \emph{compositional} 
or \emph{modular} verification. The essential idea is to verify a program ``in bits and
pieces", analyzing only a single component at a time, along with an abstraction of the environment of the component (i.e., the rest of the program). The foundations of compositional methods were created by Owicki and Gries \cite{owicki-gries76} and Lamport \cite{lamport77}. These methods are based on a simple proof rule; however, formulating the right combination of assertions for the proof rule can be a difficult and frustrating task. Also, it is not clear that doing so reduces the manual proof effort in any appreciable way, as Lamport points out in~\cite{lamport-1997}. On the other hand, for \emph{fully automated} proof methods such as model checking or static analysis, there is indeed much to be gained by a divide-and-conquer strategy, as the state space of a single component is much smaller than that of the full program. 

In this article, we focus on the simplest, but fundamental verification task: constructing an inductive program invariant. We show how the construction of a \emph{compositional} inductive invariant may be formulated as a simultaneous least fixpoint calculation. That paves the way for a variety of simplifications and generalizations. We show how the fixpoint can be computed in parallel, how the fixpoint computation may be simplified drastically by analyzing the \emph{local symmetries} of a process network, and how it may be generalized through the use of \emph{local abstraction}. As much of this exposition is based on published work, in this article we keep a light touch on the theory, emphasizing instead the intuition which lies behind the theoretical ideas. 

To illustrate the many aspects of the theory, we use a running example of a Dining Philosophers' protocol. This is chosen for two reasons: it is particularly amenable to compositional reasoning, and the protocol is flexible enough to operate on arbitrary networks, making it easy to illustrate the effects of localized symmetry and local abstraction, and their influence on parametric reasoning.

\section{Split Invariance}
\comment{\outline{
-- quantified invariants for parametric systems -- turn into split invariants -- general form -- simplification to fixpoint -- connections to O-G (give proofs) -- connections to R-G (give proofs) -- sequential complexity -- parallel complexity -- illustration with mutex protocol -- pairwise and neighborly invariants
}}

Methods for program verification are based on two fundamental concepts: that of
\emph{inductive invariance} and \emph{ranking}. An \emph{inductively invariant}
set is closed under program transitions and includes 
all  reachable program states. A \emph{ranking} function which decreases for every non-goal state shows that the program always progresses towards a goal.  The strongest (smallest)
inductive invariant set is the set of reachable states. The standard model
checking strategy -- without abstraction -- is to compute the set of reachable
states in order to show that a property is invariant (i.e., it includes all reachable states). The reachability
calculation can be prohibitively expensive due to state explosion: for instance,
the model-checker SPIN~\cite{spin} runs out of space checking the exclusion property for
approximately 10 Dining Philosophers on a ring. The divide-and-conquer approach
to invariance, which we discuss in this paper,  is to calculate an inductive
invariant which is made up of a number of local invariant pieces, one per
process. A rather straightforward implementation of this calculation  verifies
the exclusion property for 3000 philosophers in about 1 second. In this section,
we develop the basic theory behind the compositional reasoning
approach. Subsequent sections explore connections to symmetry, abstraction, and
parametric verification, as well as some of the limitations of compositional reasoning. 

\paragraph{A Note on Notation.} We use the notation developed by Dijkstra and
Scholten in~\cite{dijkstra-scholten90}. Validity of a formula $\phi$ is denoted
$[\phi]$. (We usually omit the variables on which $\phi$ depends when that can be determined from context.) Existential quantification of a set of variables $V$ is denoted $(\exists V: \phi)$. Thus, if $f$ and $g$ are formulas representing sets, $[f \Implies g]$ denotes the property that the set $f$ is a subset of the set $g$. The advantage of this notation is in its succinctness and clarity, as will be seen  in the rest of the paper. 

\subsection{Basics}
A \emph{program} is defined symbolically as a tuple $(V,I,T)$, where $V$ is a non-empty set of typed \emph{variables}, $I$ is a Boolean-valued \emph{initial} assertion defined over $V$, and $T(V,V')$ is a Boolean-valued \emph{transition} relation, defined over $V$ and an isomorphic copy, $V'$. For each variable $x$ in $V$, its copy, $x'$, denotes the value of $x$ in a successor state. 

A program defines a \emph{transition system}, represented by the tuple $(S,S^0,R)$, as follows. The set of \emph{states}, $S$, is the set of all (type-consistent) valuations to the variables $V$; the subset of \emph{initial states}, $S^0$, is those states which satisfy the initial condition $I$; and a pair of states, $(s,t)$, is in the \emph{transition relation} $R$ if $T(s,t)$ holds. 

\begin{figure}
\begin{center}
\includegraphics[scale=0.75]{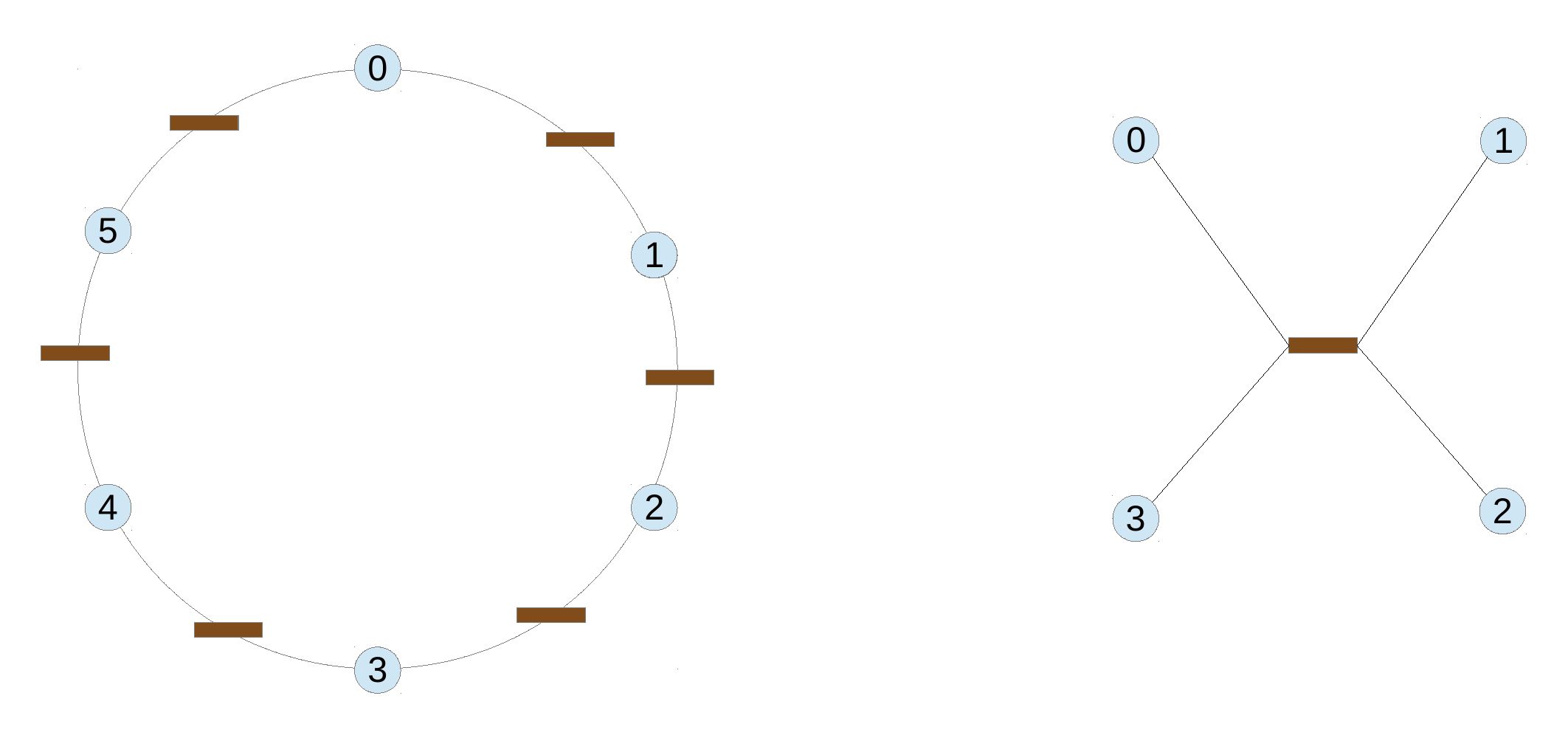}
\end{center}
\caption{Network Structure: circles represent nodes and processes, rectangles represent edges and shared state. In these networks, connectivity is bidirectional.}
\label{fig:networks}
\end{figure}

To make the role of locality clear, we work with programs that are structured as a \emph{process network}.  The network is a graph structure where nodes are labeled with programs (also called processes) and edges are labeled with shared state. Formally, the graph underlying the network is a tuple $(N,E,C)$, where $N$ is a set of \emph{nodes}, $E$ is a set of \emph{edges}, and $C$ is a \emph{connectivity} relation, a subset of $(N\times E) \Union (E \times N)$. The structure of a ring network and that of a star network is shown in Figure \ref{fig:networks}.

The \emph{neighborhood} of a node is the set of edges that are connected to it. I.e., for a node $n$, the set of edges $\{e \st (e,n) \in C \Or (n,e) \in C\}$ forms its neighborhood. Two nodes are \emph{adjacent} if their neighborhoods have an edge in common. A node $m$ \emph{points-to} a node $n$ if there is an edge $e$ in the neighborhood of $n$ such that $(m,e)$ is in $C$. 

An \emph{assignment} is a mapping of programs to the nodes of the network and of state variables to the edges. This must be done so that the program which is mapped to node $n$ only accesses, other than its internal variables, only those external variables which are mapped to edges in the neighborhood of $n$. We also require that the process only reads those variables on edges $e$ such that $(e,n)$ is a connection, and only writes those variables on edges $e$ such that $(n,e)$ is a connection. The semantics of the process network is formally defined as a program $P=(V,I,T)$, where

\begin{itemize}
\item	$V$ is the union of all program variables $V=(\Union i: V_i)$. The variables in $V_i$ which are not mapped to network edges are  the \emph{internal variables} of process $P_i$, and are denoted by $L_i$. 

\item 	$I$ is any initial condition over the program state, whose projection on $V_i$ is $I_i$. Notationally, $[(\exists V\Diff V_i: I) \Equiv I_i]$.

\item 	$T$ is the transition condition which enforces asynchronous interleaving. Formally, $[T \Equiv (\Or i: T_i \AAnd \unch(V\Diff V_i)]$. I.e., $T$ is the disjunction of the individual process transitions, under the constraint that the transition of process $P_i$ leaves all variables other than those of $V_i$ unchanged. For a simpler notation, we adopt the convention that $T_i$ is defined so that it leaves other variables unchanged.Then  $T$ can be written as $(\Or i: T_i)$. 
\end{itemize}

The set of \emph{reachable states} of a program $P=(V,I,T)$ is denoted by $\Reach(P)$, and is defined as the least fixpoint expression $(\mu Z: I \Or \SP(T,Z))$.  The fixpoint expression denotes the least (smallest, strongest) set $Z$ which satisfies the fixpoint constraint $[Z \Equiv I \Or \SP(T,Z)]$. The \emph{strongest post-condition} operator is denoted $\SP$; for a transition relation $T$ and a set of states $Z$, the expression $\SP(T,Z)$ denotes the immediate successors of states in $Z$ due to transitions in $T$. Formally, $\SP(T,Z) = \{t \st (\exists s: Z(s) \AAnd T(s,t)\}$.

A set of states (or assertion) $\varphi$ is an \emph{invariant} for the program if it is true for all reachable states, i.e., $[\Reach(P) \Implies \varphi]$. An assertion $\varphi$ is \emph{inductively invariant} if it is invariant, and also closed under program transitions. These conditions can be succinctly expressed by (1) (initiality) $[I \Implies \varphi]$, and (2) (step) $[\SP(T,\varphi) \Implies \varphi]$. Inductive invariance forms the basis of proof rules for program correctness.

\subsection{Generalized Dining Philosophers' Protocol} 
We will use a Dining Philosophers' protocol as a running example. The protocol consists  of a number of similar processes operating on an arbitrary
network. Every edge on the network models a shared ``fork''. The  edge between nodes $i$ and $j$ is called $f_{ij}$. Its value can be one of $\{i,j,\bot\}$.  Node $i$ is said to \emph{own} the fork $f_{ij}$ if $f_{ij}=i$; node $j$ owns this fork if $f_{ij}=j$; and the fork is available if $f_{ij}=\bot$. 

The process at node $i$ goes through the following internal states: $T$ (thinking); $H$
(hungry); $E$ (eating); and $R$ (release), which are the values of its internal variable, $L$. The local state of a node also includes the state of each of its adjacent edges (i.e., forks).  Let $nbr(i,j)$ be a predicate true for nodes $i,j$ if they share an edge. The transitions for a process are defined in guarded command notation as follows. 
\begin{itemize}
\item A transition from $T$ to $H$ is always enabled. I.e., $(L=T) \guards L := H$
\item In state $H$, the process acquires forks, but may also choose to release them
  \begin{itemize}
  \item (acquire fork) $(L=H) \AAnd nbr(i,j) \AAnd f_{ij} = \bot \guards f_{ij} \becomes i$,
  \item (release fork) $(L=H) \AAnd nbr(i,j) \AAnd f_{ij} = i \guards f_{ij} \becomes \bot$, and 
  \item (to-eat) $(L=H) \AAnd (\forall j: nbr(i,j): f_{ij}=i) \guards L \becomes E$. 
  \end{itemize}
\item A transition from $E$ to $R$ is always enabled. I.e., $(L=E) \guards L := R$.
\item In state $R$, the process releases its owned forks. 
  \begin{itemize}
  \item (release fork) $(L=R) \AAnd nbr(i,j) \AAnd f_{ij} = i \guards f_{ij} \becomes \bot$ 
  \item (to-think) $(L=R) \AAnd (\forall j: nbr(i,j): f_{ij} \neq i) \guards L := T$ 
  \end{itemize}
\end{itemize}
The initial state of the system is one where all processes are in internal state
$T$ and all forks are available (i.e., have value $\bot$). 
The desired safety property is that there is no reachable global state where two neighboring processes are in
the eating state $E$.

\subsection{Split Invariance}

An inductive invariant, in general, depends on all program variables; i.e., it
can express arbitrary constraints among the program variables. The divide and
conquer principle suggests that one should break up an invariance assertion 
into a number of assertions which are limited in scope, each depends only on the variables of a single process. Hence,
we define a \emph{split assertion} $\theta$ to be a conjunction, written $(\AAnd
i: \theta_i)$, of a number of local assertions $\{\theta_i\}$. The $i$'th
assertion, $\theta_i(V_i)$, is a function only of the variables  of process $P_i$; i.e., its internal variables, and those assigned to the neighborhood of node $i$.  

\begin{figure}
\begin{center}
\includegraphics[scale=0.5]{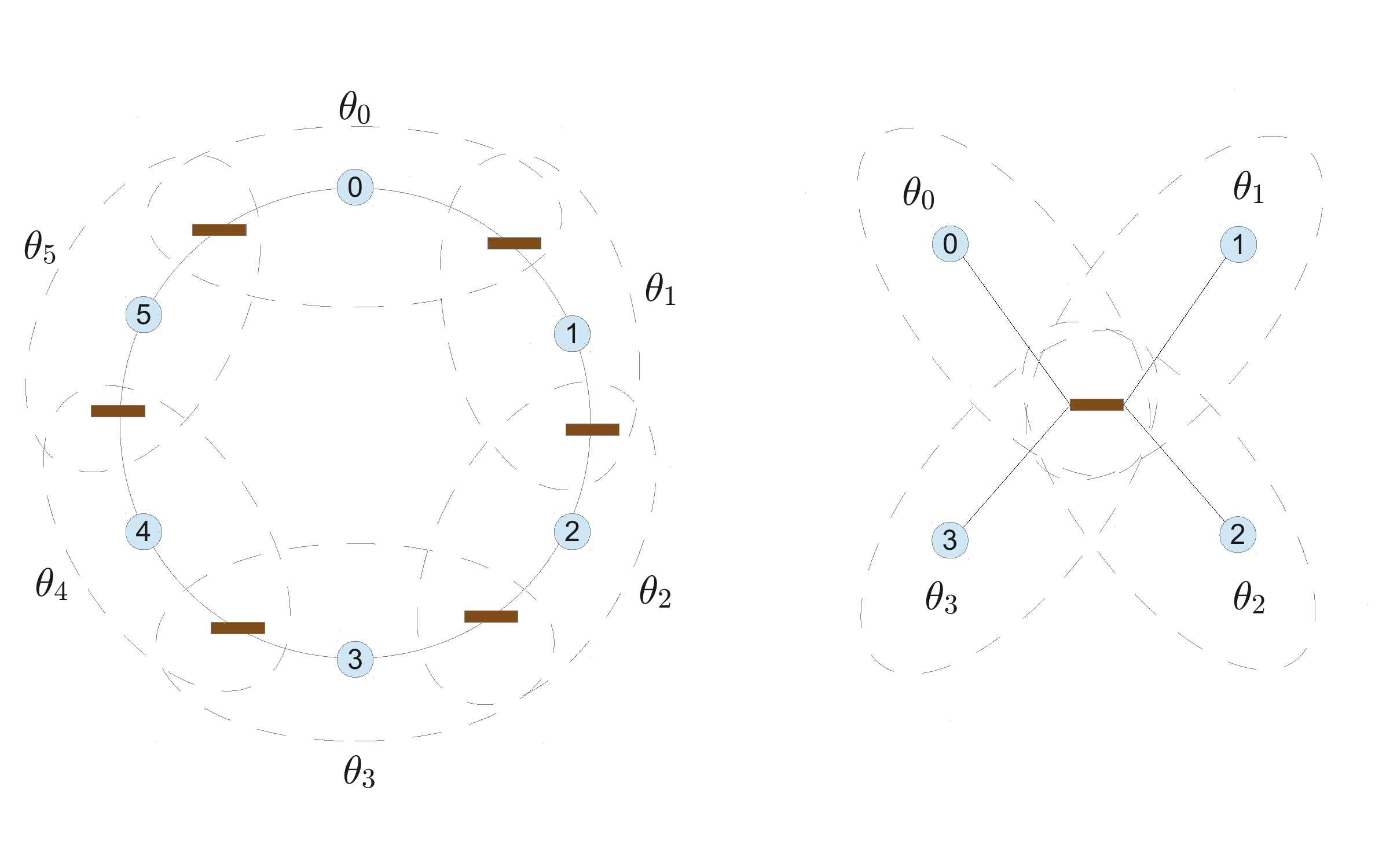}
\end{center}
\caption{Split Invariance. A dotted ellipse shows the scope of a term of the
  split invariant. }
\label{fig:splitinv}
\end{figure}

Figure \ref{fig:splitinv} gives
a pictorial view of a split invariant for the ring and star networks, showing
the scope of each of its terms. The terms for adjacent nodes have the shared variables in
common; this sets up (weak) constraints between the invariant states of those
processes.

We now consider the conditions for a split assertion to be a global inductive invariant. Examining the 
 initiality and step conditions, one notices that, as the split assertion is conjunctive and $\SP$ distributes over the disjunction of transition relations, those conditions simplify to the equivalent set of constraints given below.
\begin{align}
& [I \Implies \theta_i] \label{eq:split-init}  \\ 
& [\SP(T_i, \theta) \Implies \theta_i] \label{eq:split-step}  \\
& [\SP(T_j, \theta) \Implies \theta_i], \mbox{ for all } j \mbox{ which point to } i  \label{eq:split-intf}
\end{align}

In the last constraint, nodes $j$ which do not point to $i$ are not considered,
as any action in $T_j$ must leave the state of $V_i$ unchanged since there are
no variables in common. 

\comment{ As $\theta_i$ is defined only in terms of $V_i$, one can quantify out all other variables on the left-hand side, to obtain the equivalent set of equations: 
\begin{align}
& [(\exists V\Diff V_i: I) \Implies \theta_i] \label{eq:split-init-2}  \\ 
& [(\exists V\Diff V_i:\SP(T_i, \theta)) \Implies \theta_i] \label{eq:split-step-2}  \\
& [(\exists V\Diff V_i:\SP(T_j, \theta)) \Implies \theta_i], \mbox{ for } j \mbox{ points-to } i  \label{eq:split-intf-2}
\end{align}
}

The form of these constraints is remarkably similar to the ``assume-guarantee'' or Owicki-Gries rules for compositional reasoning, which can be stated as follows.  
\begin{align}
& [I \Implies \theta_i] \label{eq:ag-init}  \\ 
& [\SP(T_i, \theta_i) \Implies \theta_i] \label{eq:ag-step}  \\
& [\SP(T_j, \theta_i \AAnd \theta_j) \Implies \theta_i], \mbox{ for } j \mbox{ points-to } i  \label{eq:ag-intf}
\end{align}

The first two constraints show that $\theta_i$ is an invariant of process $P_i$
by itself. The third constraint is Owicki and Gries' \emph{non-interference}
condition: a transition by any other process from a state satisfying both
process' invariants, preserves $\theta_i$. The closeness of the connection
between the two formulations is shown by the following theorem. 

\theorem{\label{thm:ag-split}
Every solution to the assume-guarantee constraints is a split inductive
invariant. Moreover, in a network where all processes refer to common shared
state (such as the star network in Figure \ref{fig:networks}), the strongest solutions of the two sets of constraints are identical.}

For computational purposes, we are interested in the strongest solutions, as they correspond to least fixpoints. We will use the assume-guarantee form from now on, as it is is simpler to manipulate.  As $\theta_i$ is defined only in terms of $V_i$,  projecting the  left-hand sides of the implications \mref{ag-init}-\mref{ag-intf} on $V_i$ gives an equivalent set of constraints:
\begin{align}
& [(\exists V\Diff V_i: I) \Implies \theta_i] \label{eq:ag-init-2}  \\ 
& [(\exists V\Diff V_i: \SP(T_i, \theta_i)) \Implies \theta_i] \label{eq:ag-step-2}  \\
& [(\exists V\Diff V_i: \SP(T_j, \theta_i \AAnd \theta_j)) \Implies \theta_i], \mbox{ for } j \mbox{ points-to } i  \label{eq:ag-intf-2}
\end{align}
These constraints can be reworked into the simultaneous pre-fixpoint form (cf.~\cite{flanagan-qadeer-03,namjoshi-07a})
\begin{equation}
[F_i(\theta) \Implies \theta_i] \label{eq:prefix}
\end{equation}
where $F_i$ is the disjunction of the left-hand-sides of equations \mref{ag-init-2}-\mref{ag-intf-2}. By the monotonicity of $\SP$, the function $F_i$ is monotonic in $\theta$, considered now as a vector of local assertions, $(\theta_1,\ldots,\theta_N)$, ordered by point-wise implication. By the Knaster-Tarski theorem, there is a least fixpoint, which defines the strongest compositional invariant. It  can be computed by the standard iteration shown in Figure \ref{fig:computation}. 

\begin{figure}
\begin{center}
\begin{verbatim}
var theta, new_theta: prediate array
// initialize
for i := 1 to N do new_theta[i] := emptyset done;
// compute until fixpoint
repeat
  theta := new_theta;
  for i := 1 to N do new_theta[i] := F(i,theta) done;
until (theta = new_theta)
\end{verbatim}
\end{center}
\caption{Computing the Strongest Split Invariant.}
\label{fig:computation}
\end{figure}

The computation takes time polynomial in $N$, the number of processes in the
network -- a rough bound is $O(N^2 * L^3 * D)$, where $L$ is the size of the
local state space of a process, and $D$ is the maximum degree of the
network. This  computation produces the split invariant for the Dining
Philosophers on a 3000 node ring in about 1 second. A number of experimental
results can be found in~\cite{cohen-namjoshi07} and
in~\cite{cohen-namjoshi-saar10b}.

\subsection{Completeness}
\comment{\outline{
-- auxiliary variables -- locks, single and nested ''last'' variable (Vineet work) -- how to infer auxiliary variables -- connections with other strategies, e.g., learning. -- liveness and fairness properties
}}

The split invariance formulation is, in general, incomplete. That is, it is not
always possible to prove a program invariant by exhibiting a stronger split
invariant. In part, this is indicated by the  complexity bounds: the split
invariance calculation is polynomial in $N$, while the  invariance checking
problem is PSPACE-complete in $N$. However, this reasoning depends on whether
PSPACE=P. An direct, unconditional proof is given by the simple, shared-memory mutual exclusion program below. Every process of the program goes through states T (thinking), H (hungry), and E (eating). The desired invariance property is that no two processes are in state E together.

\begin{verbatim}
var x: boolean; initially true

process P(i): 
var l: {T,H,E}; initially T

while (true) {
  T: skip;
  H: <x -> x := false>  // atomic test-and-set
  E: x := true
}
\end{verbatim}

The fixpoint calculation  produces the split invariant $\theta$ where $\theta_i=\true$, for all $i$. This invariant clearly does not suffice to show mutual exclusion. As recognized by Owicki-Gries and Lamport, one can strengthen a split invariant by introducing auxiliary global variables which record part of the history of the computation. Intuitively, the auxiliary state helps to tighten the constraints between the $\theta$ components, as a pair of local invariant states must agree on the shared auxiliary state. For this example, it suffices to introduce an auxiliary global variable, \verb|last|, which records the last process to enter its $E$ state.

\begin{verbatim}
var x: boolean; initially true
var last: 0..N; initially 0

process P(i): 
var l: {T,H,E}; initially T

while (true) {
  T: skip;
  H: <x -> x := false; last := i>  // atomic test-and-set
  E: x := true
}
\end{verbatim}

 In the fixpoint, the $i$'th component $\theta_i$ is given by $(E_i \Equiv (\Not
 x \AAnd last=i))$. This suffices for mutual exclusion -- if distinct processes
 $P_m$ and $P_n$ are both in state $E$, then $last$ must simultaneously be equal
 to $m$ and to $n$, which is impossible. 

An important question in automated compositional model checking is, therefore,
the development of heuristics for discovering appropriate auxiliary
variables. For split invariance, one such heuristic is developed
in~\cite{cohen-namjoshi07}. It is based on a method which analyzes
counter-examples to expose aspects of the internal state of a process as an
auxiliary global predicate. The method is complete for finite-state processes:
in the worst case, all of the internal process state is exposed as shared state,
which implies that the fixpoint computation turns into reachability on the
global state space. The intuition is that for many protocols, it is unnecessary
to go to this extreme in order to obtain a strong enough invariant. This
intuition can be corroborated by experiments such as those
in~\cite{cohen-namjoshi07}. In the automaton-learning approach to compositional 
verification~\cite{cobleigh-et-al-2003}, the auxiliary state is represented
by the states of the learned automata.

\section{Local Symmetries}
\comment{\outline{
-- note symmetries in mutex. -- what are the minimum symmetries needed? -- local symmetries, groupoids, rings and torus -- collapse -- reductions in mutex and other protocol complexities. 
}}

Many concurrent programs have inherent symmetries. For instance, the mutual exclusion protocol of the previous section has a fully symmetric state space, while the Dining Philosopher's protocol when  run on a ring network has state space which is invariant under ring rotations. Symmetries can be used to reduce the state space that must be explored for model checking, as shown in the pioneering work in~\cite{clarke-filkorn-jha93,emerson-sistla93,ip-dill96}. Global symmetry reduction, however, works well only for fully symmetric state spaces, where it can result in an exponential reduction. For a number of other regular networks, such as the ring, torus, hypercube, and mesh networks, there is not enough global symmetry: the state space reductions are usually    at most linear. 

This earlier work on symmetry is connected to model checking on the full state space. What is the corresponding notion for compositional methods? It is the nature of compositional reasoning that the invariant of a process depends only on that of its neighbors. Thus, intuition suggests that it should suffice for the network to have enough \emph{local} symmetry. For example, any two nodes on a ring network are locally symmetric: each has a single left neighbor and a single right neighbor. Torus, mesh and hypercube networks also have similar local symmetries. 

Technically, the notion of local symmetry is best described by a
\emph{groupoid}~\cite{weinstein-1996}. A groupoid is a weaker object than a
group (which is used to describe global symmetries), but has many similar
properties. We use a specific groupoid, developed
in~\cite{golubitsky-stewart-2006}, which defines the local symmetries of a
network. The elements of the network groupoid are triples of the form
$(m,\beta,n)$, where $m$ and $n$ are nodes of the network, and $\beta$ is an
isomorphism on their neighborhoods which preserves the direction of connectivity. (I.e., $(m,e)$ is a connection if, and only if $(n,\beta(e))$ is a connection and, similarly, $(e,m)$ is a connection if, and only if, $(\beta(e),n)$ is a connection.) We call such a triple a \emph{local symmetry}. Local symmetries have group-like properties: 
\begin{itemize}
\item The composition of local symmetries is a symmetry: if $(m,\beta,n)$ and $(n,\delta,k)$ are symmetries, so is $(m,\delta\beta,k)$
\item The symmetry $(m,id,m)$ is the identity of the composition
\item If $(m,\beta,n)$ is a symmetry, the symmetry $(n,\beta^{-1},m)$ is its inverse.
\end{itemize}
The set of all local symmetries forms the \emph{network groupoid}. 

One may reasonably conjecture that nodes which are locally symmetric have
isomorphic compositional invariants. I.e., if $(m,\beta,n)$ is a symmetry, then
$\theta_m$ and $\theta_n$ are isomorphic up to $\beta$. This is, however, not
true in general. The reason is that the compositional invariant computed at
nodes $m$ and $n$ depends on the invariants computed at adjacent nodes, and
those must be symmetric as well. Thus, one is led to a notion of recursive similarity, called \emph{balance}~\cite{golubitsky-stewart-2006}. This has a co-inductive form like that of bisimulation. 

A balance relation $B$ is a sub-groupoid of the network groupoid, with the
following property: if $(m,\beta,n)$ is in $B$, and node $k$ points to $m$, there is a node $l$ which points to $n$ and an isomorphism $\delta$, such that $(k,\delta,l)$ is in $B$. Moreover, $\beta$ and $\delta$ must agree on the mapping of edges which are common to the neighborhoods of $m$ and $k$. 

The utility of the balance relation is given by the following theorems. 

\theorem{\label{thm:balance}
(From~\cite{namjoshi-trefler-vmcai-2012}) 
\begin{enumerate}
\item 	If $G$ is a group of automorphisms for the network, then the set $\{(m,\beta,n) \st \beta \in G \AAnd \beta(m)=n \}$ is a balance relation.
\item 	Let $\theta$ be the strongest compositional invariant for a network. 
If $(m,\beta,n)$ is in a balance relation, then $[\theta_n \Equiv \dia{\beta}\theta_m]$.
\end{enumerate}}

Informally, the first result shows that the global symmetry group induces
balanced local symmetries; this is a quick way of determining a balance relation
for a network. The second shows that the local invariants for a pair of balanced
nodes  are isomorphic. Here,  $\dia{\beta}$ is a pre-image operator that maps states over $V_m$ to states over $V_n$ using $\beta$ to relate the values of corresponding edges. 

\paragraph{Local Symmetry Reduction.}
This theorem points the way to symmetry reduction for compositional methods. The idea is to compute fixpoint components only for representatives of local symmetry classes. The group-like properties ensure that for any groupoid, its \emph{orbit relation}, defined as $m \sim n$  if there is $\beta$ such that $(m,\beta,n)$ is in the groupoid, is an equivalence. For a ring network, it suffices to compute a single component, rather than all $N$ components! The calculation is thus independent of the size of the  network. This has interesting consequences for parametric proofs, as explained in the next section.

\section{Local Abstractions}
\comment{\outline{
-- why needed? Dining phils -- collapse to generic process -- generalization to parametric case -- 
}}

The  symmetry reductions described in the previous section are applicable
to several networks which have only a small amount of global symmetry. Still,
protocols have other local symmetries which cannot be captured by this definition. For
instance, consider the Dining Philosophers' protocol on an arbitrary
network. Every node operates in a roughly similar fashion, attempting to own all
of its forks before entering the eating state. However, nodes with differing numbers of adjacent edges cannot be locally symmetric -- there can be no isomorphism between their neighborhoods. In fact, a network may be so irregular as to have only the trivial symmetry groupoid. 

In order to be able to represent these other symmetries, we must abstract away from the structural differences between nodes. It suffices to define an abstraction function over the local state of a node. As explained in more detail in~\cite{namjoshi-trefler-vmcai-2013}, a \emph{local abstraction} is formally specified by defining for each node $m$ an abstract domain, $D_m$, and a total abstraction function, $\alpha_m$, which maps local states of $P_m$ to elements of $D_m$. This induces a Galois connection on subsets, which we also refer to as $(\alpha_m,\gamma_m)$: $\alpha_m(X) = \{\alpha_m(x) \st x \in
X\}$, and $\gamma_m(A) = \{x \st \alpha_m(x) \in A\}$. 

We must adjust the fixpoint computation to operate at the abstract state level. 
The abstract set of initial states, $\abs{I}_m$ is given by $\alpha_m(I_m)$. 
The abstract step transition, $\abs{T}_m$, is obtained by standard existential abstraction: there is a transition from (abstract) state $a$ to (abstract) state $b$ if there exist local states $x, y$ 
such that $\alpha_m(x)=a$, $\alpha_m(y)=b$, and $T_m(x,y)$ holds.  
An abstract transition $(a,b)$ for node $m$ is the result of interference by a
transition of node $k$ from $\theta_k$ if the following holds. 
\begin{equation}
  \label{eq:absintf}
  (\exists s,t: \alpha_m(s[m])=a \AAnd \alpha_m(t[m])=b  \AAnd T_k(s,t) \AAnd \alpha_k(s[k]) \in \theta_k)
\end{equation}

\begin{figure}
\begin{center}
\scalebox{0.50}{\includegraphics{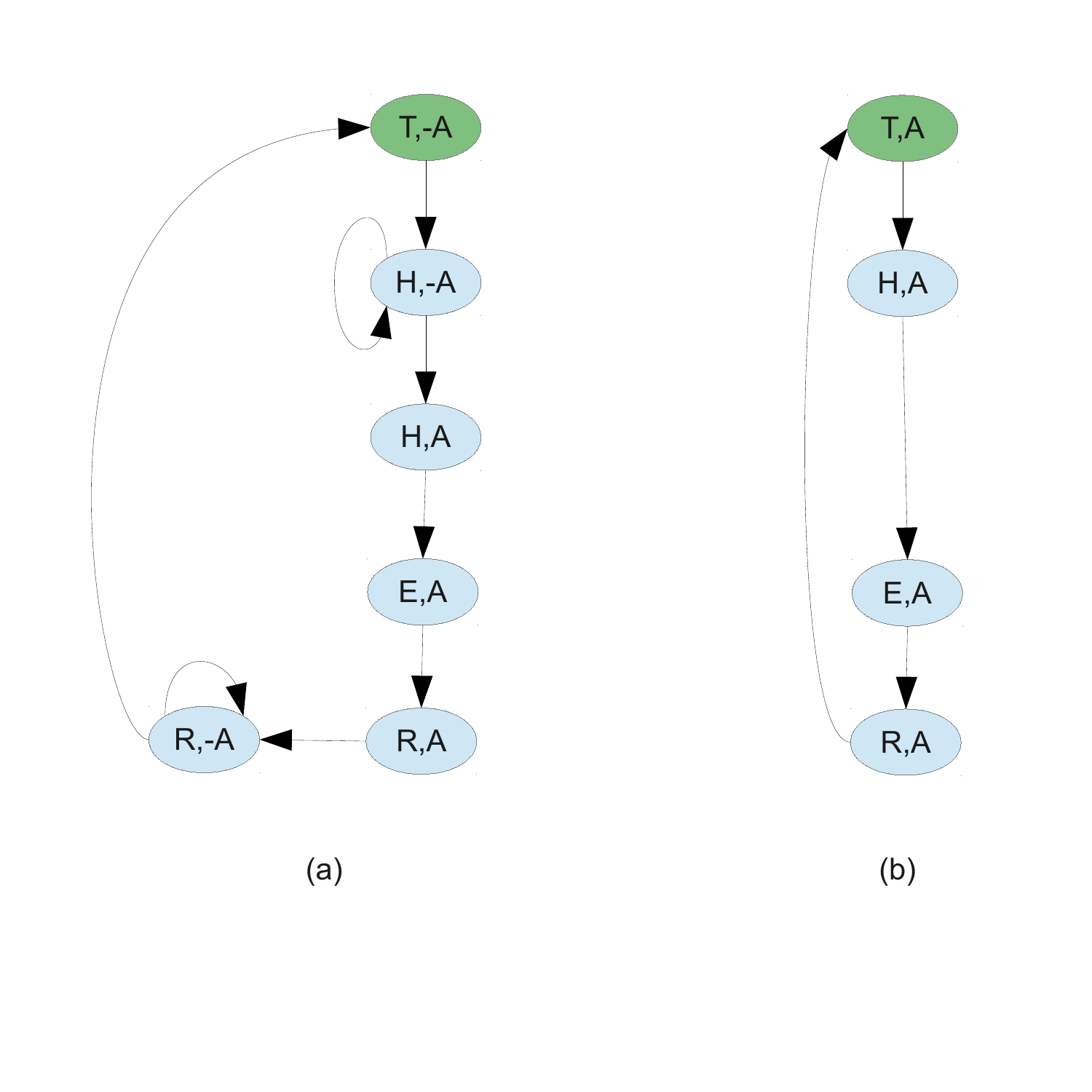}}
\caption{(From~\cite{namjoshi-trefler-vmcai-2013}) Abstract State Transitions (a) for non-isolated nodes and (b)
  for an isolated node. The notation ``$-\pA$'' indicates the negation of
  $\pA$. Green/dark states are initial. }
\label{fig:abs-dp}
\end{center}
\end{figure}

For the Dining Philosophers' protocol, such a function can be defined through a predicate, $\pA$, which is true at a local state if, and only if, the node owns all forks in that state. The abstract state of a node is now a pair $(l,a)$ where $l$ is its internal state (one of $T,H,E,R$) and $a$ is the value of the predicate $\pA$. With this definition, the abstract fixpoint calculation produces the local invariant shown as a transition graph in Figure \ref{fig:abs-dp}. This graph shows that the abstract invariant for node $m$ implies that $(E_m \Implies \pA_m)$. Concretizing this term, one obtains that the concrete invariant implies that  if $E_m$ is true, then node $m$ owns all of its forks. This, in turn, implies the exclusion property, as adjacent nodes $m$ and $n$ cannot both own the common fork $f_{mn}$ in the same global state.

There are two features to note of this transition graph. First, all interference
transitions are self-loops -- i.e., the actions of neighboring processes do not
change the abstract state of a process. This is due to the protocol: the action
of a neighbor cannot cause a process to own a fork, or to give up one that it
owns.  Second, all nodes in any network fall into one of the two classes which
are shown, in terms of their abstract compositional invariant. It follows that
the concretized compositional invariant holds in a parametric sense: i.e., over
\emph{all} nodes of \emph{all} networks. 

This connection between compositional reasoning and parametric proofs is not
entirely unexpected. Parametric invariants for protocols often have the
universally quantified form ``for every node $n$ of an instance, $\theta(n)$
holds''. If the property $\theta(n)$ is restricted to the neighborhood of $n$
and holds compositionally, which is often the case, then the property is a split
invariant for every fixed-size instance. The application of abstraction and
symmetry serves to turn this connection around:  computing a compositional
invariant on an (abstract) instance induces  a parametric invariant. 

The following theorem shows that this is a complete method -- but it is not automatic, as it requires the choice of a proper abstraction. The abstraction in the theorem can be chosen so that every pair of nodes is locally symmetric in terms of its abstract state space, and the cross-node interference is benign, as in the illustration above.

\theorem{(From~\cite{namjoshi-trefler-vmcai-2013}) \label{thm:param-completeness}
	 For a parameterized family of process
  networks, any compositional invariant of the form $(\forall i: \theta_i)$, where each
   $\theta_i$ is local to process $P_i$, can be established by 
   compositional reasoning over a small abstract network.}

\section{Related Work}
The book~\cite{deRoever01} has an excellent description of the Owicki-Gries method
and other compositional methods. The fixpoint formulation for computing the
strongest split invariant is implicit in the deduction system of~\cite{flanagan-qadeer-03} and is
explicitly formulated in~\cite{namjoshi-07a}. Other closely related work on compositional verification
has been referenced in the previous sections. 

Local symmetry and balance are originally defined
in~\cite{golubitsky-stewart-2006} in a slightly different form. That paper analyzes the role of
local symmetry to prove \emph{existential} path properties of \emph{continuous}
systems; it is remarkable that those definitions also serve to analyze universal
properties of discrete systems.  it is explicitly stated

Parametric verification is undecidable in general, even if each process has a
small, fixed number of states~\cite{apt-kozen86}. A common thread running
through the various approaches to parametric verification is the intuition that
for a correct protocol, behaviors of very large instances are already present in
some form in smaller instances. Decidability results~\cite{german-sistla92,emerson-namjoshi95,emerson-kahlon00} are
 based on ``cutoff'' theorems which establish that it suffices to check all
 instances up to the cutoff size, or on well-quasi-ordering of the transition
 structure~\cite{abdulla-cerans-et-al-1996}. The method of 
 invisible invariants~\cite{pnueli-ruah-zuck01} generalizes an invariant
 computed automatically for a small instance to an inductive invariant for all
 instances. In~\cite{namjoshi-07a}, it was shown that the success of
 generalization is closely related to the invariant being a split invariant; the
 parametric analysis based on abstractions and local symmetry that is carried
 out in this paper is a further extension of those results. The ``environmental
 abstraction'' procedure~\cite{talupur06} analyzes a single process in the
 context of an approximation of the rest of the system. Although the
 approximation is developed starting with the full state
 space, there is a close similarity between the final method and compositional
 reasoning. Related procedures include~\cite{mcmillan-zuck-2011}
 and~\cite{sanchez-2012}.

\section{Conclusions}
\comment{\outline{-- nice theory -- open questions: what programs are most amenable to such reasoning? -- connections with knowledge -- strategies for synthesis? 
}}

In this article, I have attempted to show that compositional reasoning is a
topic with a rich theory and practically relevant application. There are
pleasing new connections to the new concept (in verification) of local symmetry,
and to long-established ones such as abstraction and parametric reasoning. In
this article I have chosen to focus on the simplest form of compositional reasoning, that used to construct inductive invariants, but the methods extend to general (i.e., possibly non-inductive) invariance, as well as to proofs of temporal properties under fairness assumptions~\cite{cohen-namjoshi08,cohen-namjoshi-saar10,cohen-namjoshi-saar10b}. The simultaneous fixpoint calculation lends itself to parallelization, as the individual components can be computed asynchronously so long as the computation schedule is fair~\cite{cohen-et-al-hvc2010}. It is worth noting that the theory applies to arbitrary state spaces under appropriate abstractions, as shown by the work in~\cite{gupta-popeea-rybalchenko-2011}, which  applies compositional reasoning to  C programs. 

There are many open questions. Among the major ones are the following: Why are certain protocols more amenable to compositional methods than others? (``Loose coupling'' is sometimes offered as an answer, but that term does not have a precise definition.) Can one create better methods which  compute only as much auxiliary state as is necessary for a proof? What sorts of abstractions are useful for parametric proofs? 

\comment{Finally, an intrit seems intuitively likely (cf.~\cite{namjoshi-trefler-2013}) that compositional methods work well for protocols that must operate under adversarial conditions, such as those in ad-hoc and dynamic network models. This is because any correct protocol must work relative to minimal assumptions about its neighbors, since the set of neighbors and their connectivity can change at any moment.}

\paragraph{Closing.} 
I would like to thank the referees for helpful comments on the initial draft of
this paper. The work described here would not have been possible without the varied and immensely enjoyable collaborations with my co-authors: Ariel Cohen, Yaniv Sa'ar, Lenore Zuck, and Richard Trefler. My co-authors on a survey of compositional verification, Corina P\u{a}s\u{a}reanu and Dimitra Giannakopoulou, contributed many insights into the methods. I am very glad to be able to offer this as my contribution to David Schmidt's Festschrift. It is a small return for the respect I have for his research work, for the help and advice I received from him when co-organizing VMCAI in 2006, and for many enjoyable conversations! 

My initial work on compositional reasoning was supported in part by the NSF, under award CCR 0341658. The writing of this paper was done while I was supported in part by DARPA, under agreement number FA8750-12-C-0166. The U.S. Government is authorized to reproduce and
distribute reprints for Governmental purposes notwithstanding any copyright
notation thereon. The views and conclusions contained herein are those of
the authors and should not be interpreted as necessarily representing the
official policies or endorsements, either expressed or implied, of DARPA or
the U.S. Government.

\bibliographystyle{eptcs}
\bibliography{general}
\end{document}